\documentclass[aps,prb,twocolumn,oneside,floatfix,sort&compress,
superscriptaddress,showpacs,nobalancelastpage]{revtex4-1}

\usepackage{graphicx,xcolor}
\usepackage[utf8]{inputenc}
\usepackage{amssymb,bbm}
\usepackage[reqno]{amsmath}
\usepackage{bbm}

\usepackage[normalem]{ulem}

\begin{document}
\title{Spin-spin correlations between two Kondo impurities\\ coupled to an open Hubbard chain}

\author{A. C. Tiegel}
\affiliation{Institut f\"ur Theoretische Physik, Georg-August-Universit\"at G\"ottingen, 37077 G\"ottingen, Germany}
\author{P. E. Dargel}
\affiliation{Institut f\"ur Theoretische Physik, Georg-August-Universit\"at G\"ottingen, 37077 G\"ottingen, Germany}
\author{K. A. Hallberg}
\affiliation{Instituto Balseiro, Centro At\'omico Bariloche, CNEA and CONICET, 8400 Bariloche, Argentina}
\author{H. Frahm}
\affiliation{Institut f\"ur Theoretische Physik, Leibniz Universit\"at Hannover, 30167 Hannover, Germany}
\author{T. Pruschke}
\affiliation{Institut f\"ur Theoretische Physik, Georg-August-Universit\"at G\"ottingen, 37077 G\"ottingen, Germany}
            
\date{\today}                 

\begin{abstract}
In order to study the interplay between Kondo and Ruderman-Kittel-Kasuya-Yosida (RKKY) interaction, we calculate the spin-spin correlation functions between two Kondo impurities coupled to different sites of a half-filled open Hubbard chain. Using the density-matrix renormalization group (DMRG), we re-examine the exponents for the power-law decay of the correlation function between the two impurity spins as a function of the antiferromagnetic coupling $J$, the Hubbard interaction $U$ and the distance $R$ between the impurities. The exponents for finite systems obtained in this work deviate from  previously published DMRG calculations. We furthermore show that the long-distance behavior of the exponents is the same for impurities coupled to the bulk or to both ends of the chain. We note that a universal exponent for the asymptotic behavior cannot be extracted from these finite-size systems with open boundary conditions.
\end{abstract}

\maketitle

\section{Introduction}
\label{sec: intro}
The Kondo effect\cite{Kondo1964} is one of the oldest and most studied correlation phenomena in condensed matter physics. It has regained vital interest by single-impurity setups and microscopic measurements.\cite{PhysRevLett.80.2893, Madhavan24041998, Manoharan2000}  One of the present main foci is the extension of the Kondo cloud and its experimental measurement.\cite{Affleck2001, Affleck2008, Hand2006, Prueser2011} The basic idea is that the conduction band electrons will form a highly correlated quantum state with the impurity spin and screen it.\cite{Hewson1993} This results in nonzero spin-spin correlations between the impurity spin and the conduction band electrons. The decay of these spin-spin correlations has been the subject of many theoretical studies.\cite{Sorensen1996,Barzykin1996,Costamagna2006,Holzner2009} In particular, the corresponding envelope of these correlations was found to cross over from a $1/R$ decay to a $1/R^2$ decay at the Kondo coherence length, where $R$ denotes the distance 
between the impurity and the conduction band site.\cite{PhysRevB.75.041307}

If a second impurity is added to the system, one will have two competing interactions: the Kondo effect and the Ruderman-Kittel-Kasuya-Yosida (RKKY) interaction.\cite{Ruderman1954,Kasuya1956,Yosida1957} The Kondo effect will screen the impurity spins individually, leading to vanishing spin-spin correlations between the two impurity spins. The RKKY interaction favors a magnetic interaction between the impurity spins, i.e., one observes strong correlations between the two impurity spins. The interplay between these effects in the two-impurity problem has attracted much attention.\cite{PhysRevLett.47.737, JonesVarma1988, FyeHirsch1989, AffleckLudwig1992, FyeRichard1994, PhysRevB.52.9528, Simon2005, Costamagna2008, Simonin2006} 

In this work, we want to focus on the static spin-spin correlations between two Kondo impurities. Another aspect, which is usually neglected, is the presence of correlations in the conduction band system. They can in principle further modify exponents and also introduce additional functional dependencies, such as, for example, logarithmic corrections. Therefore, we examine the system  in the presence of a finite $U$ in the conduction chain. As an analytical treatment of the two-impurity Kondo problem is not available, one has to rely on numerical solutions. We will use the density-matrix renormalization group\cite{White1992, White1993} (DMRG) to obtain the ground state and to calculate the spin-spin correlation functions. In particular, we are interested in the form and decay of these correlations. Hallberg and Egger\cite{Hallberg1997} calculated exactly these spin-spin correlations shortly after the development of the density-matrix renormalization group. From their data, they have argued that the 
correlations will show a power-law behavior in the long-distance limit $\langle \textbf{S}_I \mathbf{S}_{II}\rangle_R \propto 1/R^2$ for two Kondo impurities irrespective of the interaction $U$ in the chain.

Based on refined numerics for larger systems, which are accessible due to the great increase in computer power, we re-examine these exponents.  For $U>0$,  we show that even with this increase in computational resources, one cannot easily identify a simple $1/R^2$ power-law behavior for the spin-spin correlations between two Kondo impurities. Only if we include a $U$-dependent logarithmic correction, our data are compatible with an exponent $\alpha=2$. This investigation shows that even for system sizes accessible today, an unbiased estimation of exponents for these long-range correlations is very difficult, especially if one has to expect logarithmic corrections. It is then important to understand how well such numerical calculations can reveal exponents expected to rule the decay of correlation functions for sufficiently large distances of the impurities.

The paper is organized as follows. After presenting the model and a brief discussion of the method and its problems in Sec.~\ref{sec: model}, we present our results in Sec.~\ref{sec: corrfunc}, starting with impurities attached to the chain ends. The main results and conclusions of the paper are summarized in Sec.~\ref{sec: summary}.

\section{Model and method}
\label{sec: model}

\begin{figure}
\centering
 \resizebox{0.98\columnwidth}{!}{\includegraphics{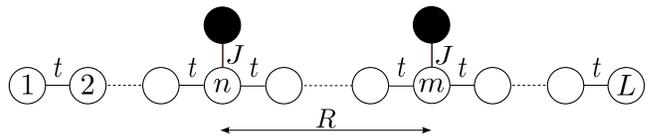}}
\caption{Two spin-$1/2$ impurities (filled circles), separated by the inter-impurity distance $R= |m-n|$, are attached to a one-dimensional Hubbard chain of length $L$.}
\label{fig: setup}
\end{figure}

We study two spin-$1/2$ Kondo impurities attached to a one-dimensional Hubbard chain. Figure\,\ref{fig: setup} shows the corresponding setup. The Hamiltonian 
\begin{align*}
H = H_c + H_{sd} 
\end{align*}
can be divided into two parts. The first contribution
\begin{align*}
H_c = - t \sum_{i,\sigma}\, \left( c_{i,\sigma}^{\dagger} c_{i+1,\sigma}^{\phantom{\dagger}} + c_{i+1,\sigma}^{\dagger} c_{i,\sigma}^{\phantom{\dagger}} \right)  + U \sum_i \, n_{i,\uparrow} n_{i,\downarrow}
\end{align*}
models the Hubbard conduction band with a Coulomb repulsion $U>0$ at each site. Here $c_{i}^{(\dagger)}$ denotes the usual fermionic  annihilation (creation) operator at site $i$ and $n_i$ represents the particle number operator. Throughout our work, we adopt open boundary conditions and the hopping parameter is set to $t=1$. For $U=0$ the conduction band $H_c$ reduces to a tight-binding model.

The coupling of the two Kondo spins to the conduction electrons is modeled by the $s$-$d$ exchange term
\begin{align*}
H_{sd} =- J_I \mathbf{S}_I \cdot \mathbf{s}_n - J_{II} \mathbf{S}_{II} \cdot \mathbf{s}_m,
\end{align*}
where $ \mathbf{S}_{I(II)}=(S^x,S^y,S^z)$ is the spin operator of the Kondo impurity which is attached to site $n(m)$ of the Hubbard chain. The operator $ \mathbf{s}_{n(m)}=(s_{n(m)}^x,s_{n(m)}^y,s_{n(m)}^z)$ is the spin operator at the conduction band site. We restrict our study to the case of antiferromagnetic coupling constants of equal strength $J_I = J_{II}= J < 0$.

\begin{figure}
\centering
\resizebox{0.98\columnwidth}{!}{\includegraphics[angle=-90]{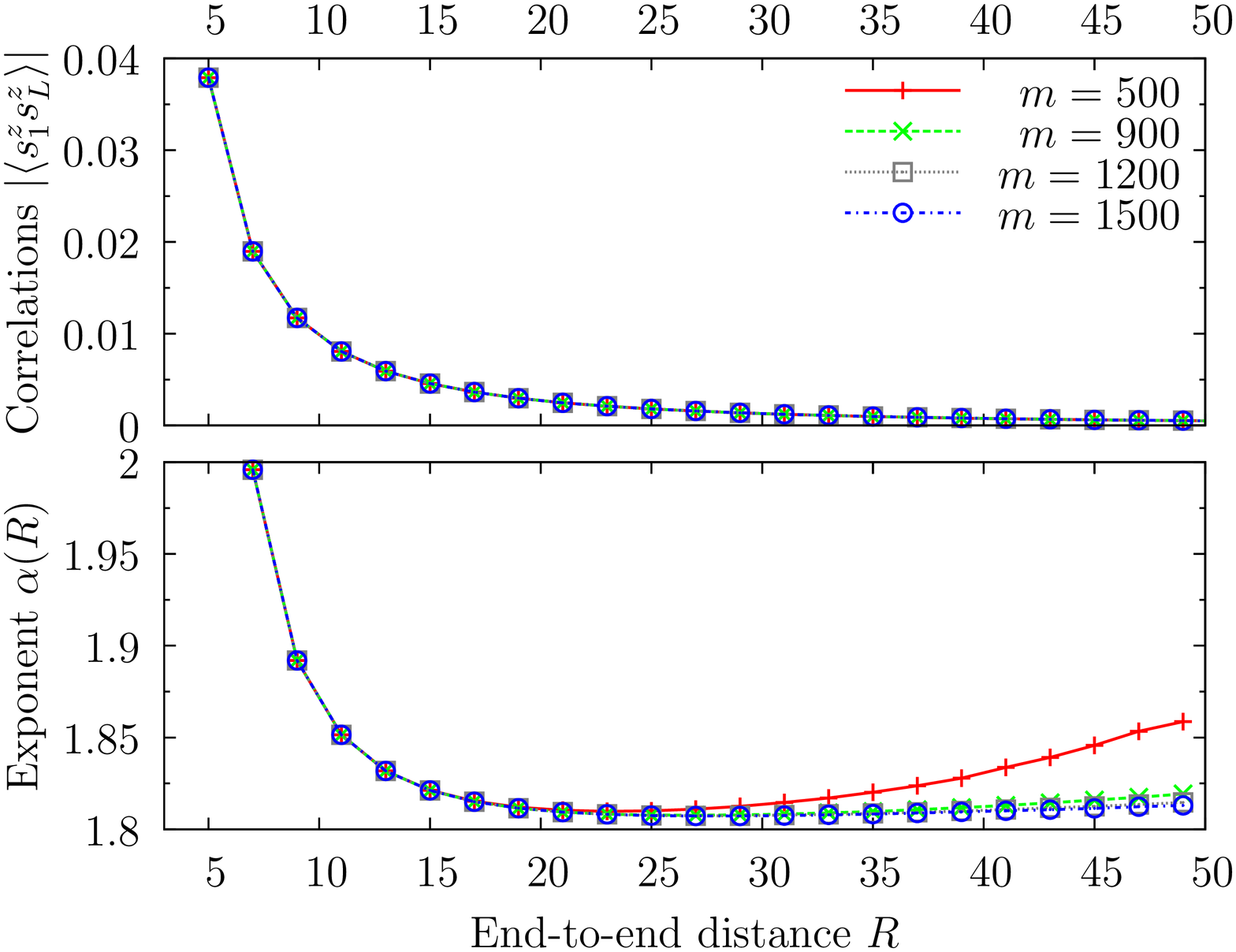}}
\caption{(Color online) Upper panel: Comparison of DMRG data for the end-to-end correlation function $\langle s^z_1 s^z_L \rangle$ of a half-filled Hubbard chain with $U=1$ on a linear scale for different numbers $m$ of kept states. Lower panel: Double-logarithmic central differences giving the exponent $\alpha(R)$ extracted from $\langle s^z_1 s^z_L \rangle$ for $U=1$ in the spirit of Eq.~\eqref{eq: cdiff}.}
\label{fig: check_U1}
\end{figure}

We use a standard density-matrix renormalization group (DMRG) algorithm\cite{Schollwock2005,Hallberg2006,Schollwoeck2011} for open boundary conditions to carry out the calculations. The setup with the two impurities attached to the ends of the chain is computationally less demanding and hence more accurate than the analysis of the bulk limit ($R \ll L/2$). Since it will turn out that the results in the long-distance or strong-coupling limit are the same for these two setups, the majority of our calculations is performed with the impurities coupled to the ends. This is still very costly because for every different inter-impurity distance, a new DMRG setup as well as a new DMRG ground-state calculation is needed. Furthermore, extracting the true long-range behavior of correlation functions within the DMRG is known to be difficult since all correlations are either long ranged or purely exponentially decaying, and even for power-law correlations this modeling only works for not too long distances (cf. Ref.~\
onlinecite{Schollwock2005}). In order to estimate the minimal number of basis states $m$ that has to be kept within the DMRG, the upper panel of Fig.~\ref{fig: check_U1} shows the end-to-end correlation function $\langle s^z_1 s^z_L \rangle$ for a half-filled Hubbard chain with $U=1$ and different values of $m$. At first sight, the DMRG data for the correlation function itself looks the same irrespective of the number of kept states. However, a more thorough analysis by means of double-logarithmic central differences (to be introduced later, see Eq.~\eqref{eq: cdiff}), shown in the lower panel of Fig.~\ref{fig: check_U1}, reveals that one needs to choose a very large number of basis states $m$ in order to obtain a reasonably converged estimate of the long-distance decay respectively the associated power law. Thus, our first conclusion is that the question of convergence very subtly depends on the property one is eventually interested in. From Fig.~\ref{fig: check_U1} we also conclude that even a reliable 
extraction of the exponent is very difficult for longer distances. Consequently, in our DMRG calculations, we typically keep $m=1300$ basis states for chains up to lengths of $L=120$.

At half-filling, we learned from exact diagonalization results for small systems of even length $L$ that the ground state is nondegenerate for $S^z_{tot} =0$, if we attach one Kondo impurity to an even site and the other one to an odd site. The inter-impurity distance is given by $R= |m-n|$. The DMRG calculations are carried out in the two Abelian U(1) symmetry sectors defined by the electron number $N$ and the $z$-component of the total spin $S^z_{tot} = S_I^z + S_{II}^z + S_c^z$. Introducing full SU(2) symmetry in the spin sector can reduce the computational effort slightly, but will not alter our main observations.

\section{Correlation functions}
\label{sec: corrfunc}

The main focus of our work is the analysis of the spatial behavior of the spin-spin correlation functions $\langle \textbf{S}_I \mathbf{S}_{II}\rangle_R$ between the two attached Kondo spins at half-filling and zero temperature. In particular, we want to determine whether for infinitely large distances $R$ a power law of the form
\begin{align}
\langle \textbf{S}_I \mathbf{S}_{II}\rangle_R \propto R^{-\alpha}\label{eq:purepowerlaw}
\end{align}  
with a constant exponent $\alpha\neq\alpha(R)$ exists.

\subsection{Impurities coupled to the ends}
\label{sec: ends}

The impurities are attached to the first and the last sites of a Hubbard chain of length $L$. The distance between the impurities therefore is $R=L-1$. The correlation function is $2k_F$ oscillatory, with $k_F$ given by the filling, e.g., $k_F=\pi/2$ for half-filling. We will concentrate on odd distances $R$ to avoid these oscillations.

\subsubsection{Non-interacting conduction chain $U=0$}
\begin{figure}
\centering
\resizebox{0.98\columnwidth}{!}{\includegraphics[angle=-90]{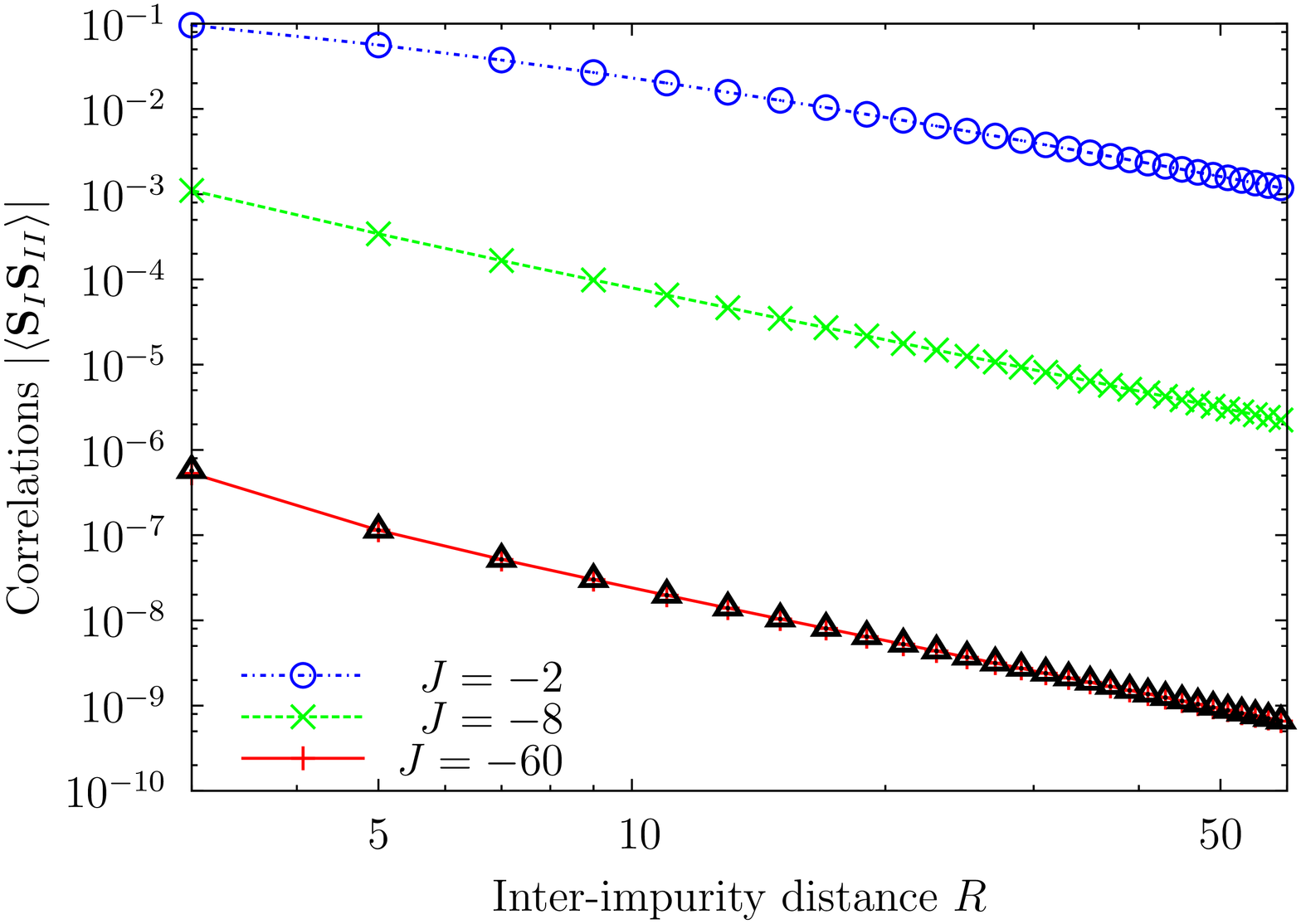}}
\caption{(Color online) DMRG data for $\langle \mathbf{S}_I \mathbf{S}_{II} \rangle_R$ on a double-logarithmic scale for $U=0$ and the impurities at the ends of the chain, including a comparison with perturbation theory (triangles) for $J=-60$.}
\label{fig: ptheory}

\end{figure}

We start by re-examining the system with a noninteracting conduction chain. For $|J| \to \infty$, the Kondo impurities form a rigidly bound singlet with the spins at the conduction band sites they are attached to, i.e., the first and the last sites. In this case, the system decouples into three parts (cf. Ref.~\onlinecite{PhysRevLett.68.1220}): the two singlets and an effective chain of length $L-2$ in-between. Thus, for strong couplings $|J| \gg 1$, there is almost no hopping between sites 1 and 2 as well as between $L-1$ and $L$. These considerations form the basis of a perturbative treatment which has originally been proposed for a single Kondo impurity coupled to one end of a noninteracting conduction band. \cite{Sorensen1996} In  Appendix~\ref{sec: ptheory}, we extend this result to the case of two impurities attached to the ends. Our DMRG results can therefore be directly compared with perturbation theory for strong couplings $|J| \gg 1$ and $U=0$. The spin-spin correlation functions at half-filling in 
second-order perturbation theory are given by
\begin{align}
& \langle \mathbf{S}_I \mathbf{S}_{II} \rangle_{R=L-1}^{(2)} = 12 \, \left(\frac{20}{9}\right)^2 \left( \frac{t}{J}\right)^4 \langle F | s^z_2  s^z_{L-1} | F \rangle \label{eq: pt_universal} \\
 &= \frac{24}{(L+1)^2} \left( \frac{20}{9}\right)^2  \left( \frac{t}{J}\right)^4 \sum_{l=1}^{L/2} \sum_{q=L/2+1}^{L} \biggl[ \sin \left( \frac{L-1}{L+1}\pi l \right) \times \nonumber \\
 &\times \sin \left( \frac{2\pi}{L+1} l \right)  \sin \left( \frac{L-1}{L+1}\pi q \right) \sin \left( \frac{2\pi}{L+1} q \right) \biggr], \label{eq: ptheory}
\end{align}
where $|F\rangle$ is the ground state of the noninteracting chain. This means that for very strong coupling the correlation function $\langle \mathbf{S}_I \mathbf{S}_{II} \rangle^{(2)}$ is effectively given by  $\langle s^z_2  s^z_{L-1} \rangle$ between the first and the last sites of the tight-binding chain in-between the two strongly bound Kondo singlets. Such a behavior can indeed be observed in Fig.~\ref{fig: ptheory}, where the DMRG data for the correlation functions are plotted on a double-logarithmic scale and are also compared to the perturbation theory for large values of $|J|$. There is very good agreement for $J=-60$ with relative deviations of less than 0.5~\%.

\begin{figure}
\centering
\resizebox{0.98\columnwidth}{!}{\includegraphics[angle=-90]{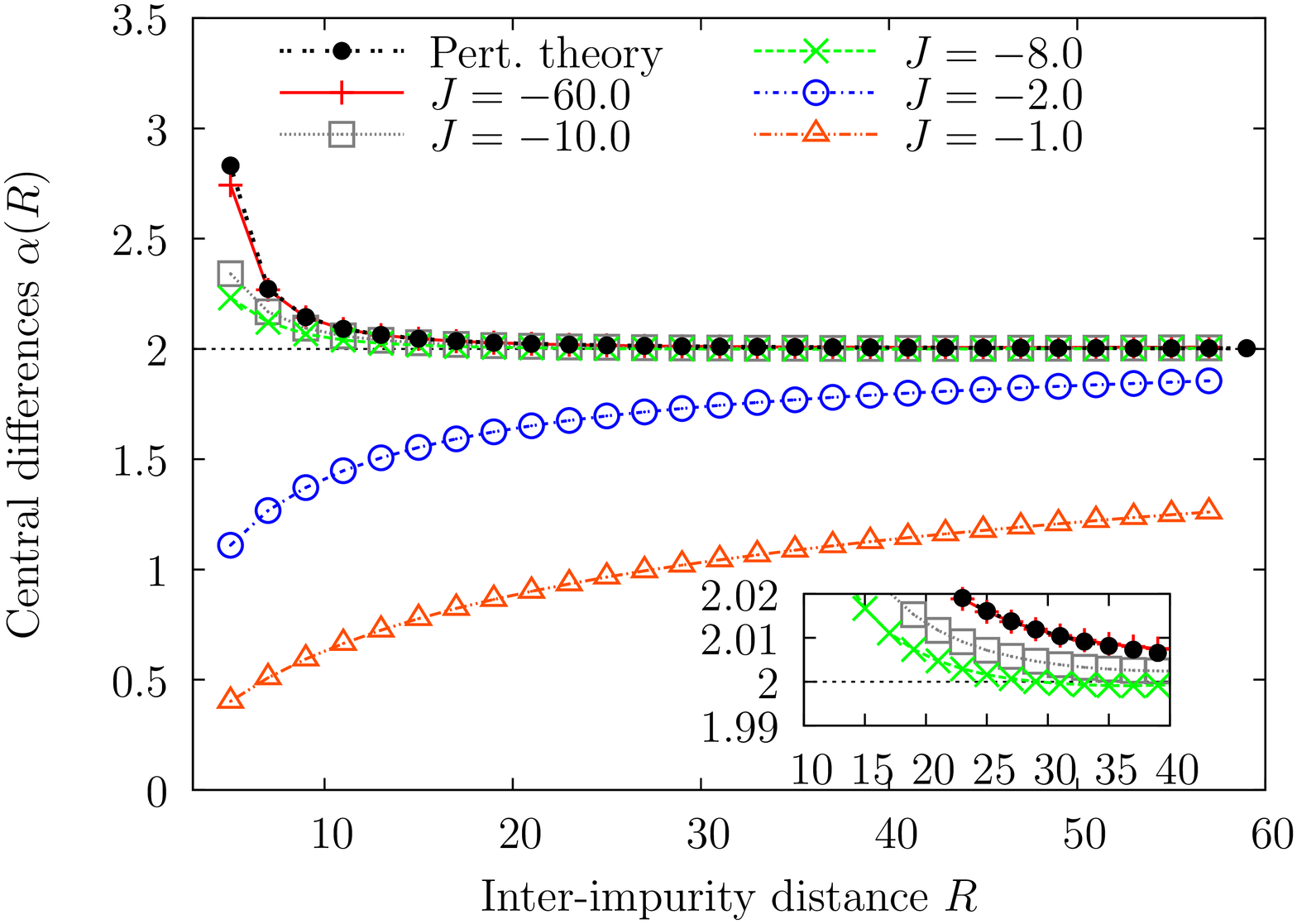}}
\caption{(Color online) Double-logarithmic central differences $\alpha(R)$ for $U=0$ and various values of $J$ including the results of the perturbation theory for $|J| \gg 1$.  }
\label{fig: alpha_U0}
\end{figure}

The results from Fig.~\ref{fig: ptheory} suggest a power-law behavior for large couplings as given by Eq.~\eqref{eq:purepowerlaw}. In order to extract the exponent and to validate the existence of a power law from our DMRG calculations, we have computed the double-logarithmic central differences
\begin{align}
\alpha(R) = \frac{\log( \langle \mathbf{S}_I \mathbf{S}_{II} \rangle_{R+2} ) - \log( \langle \mathbf{S}_I \mathbf{S}_{II}\rangle_{R-2})} {\log(R+2)-\log(R-2)} \label{eq: cdiff}
\end{align}
of the correlation functions for various couplings. They will directly give the exponent $\alpha$ if a power law $1/R^\alpha$ is assumed at a certain distance. This analysis is more accurate for the evaluation of the distance dependence $\alpha(R)$ than fitting a power law to the data of Fig.~\ref{fig: ptheory}. In Fig.~\ref{fig: alpha_U0} the central differences $\alpha(R)$ are shown for $U=0$ and several values of $J$. For strong coupling and large distances, a fast convergence $\alpha \to 2$ can be clearly observed. The asymptotic behavior is also in agreement with our perturbative results whose derivative according to Eq.~\eqref{eq: cdiff} is also shown in Fig.~\ref{fig: alpha_U0}. For small $|J|$, the accessible distances are too small to find a $1/R^2$ behavior. However, the results are in agreement with such a behavior for $R\to \infty$.

As a curiosity, we remark that there appears to be a change in behavior with decreasing $|J|$. For large $|J|$, the asymptotic exponent is consistently approached from above, while for smaller values of the coupling, the short-distance behavior is weaker than the asymptotic $1/R^2$. We interpret this as a sign of the crossover from Kondo screening at large $|J|$, i.e., outside of the scales set by the Kondo correlation length one does not have sizable influence by the spins any more, to the RKKY dominated regime at small coupling, where the correlations for small distances should rather reflect the tendency towards antiferromagnetic order with a correspondingly slow decay.\\

\begin{figure}
\centering
\resizebox{0.98\columnwidth}{!}{\includegraphics[angle=-90]{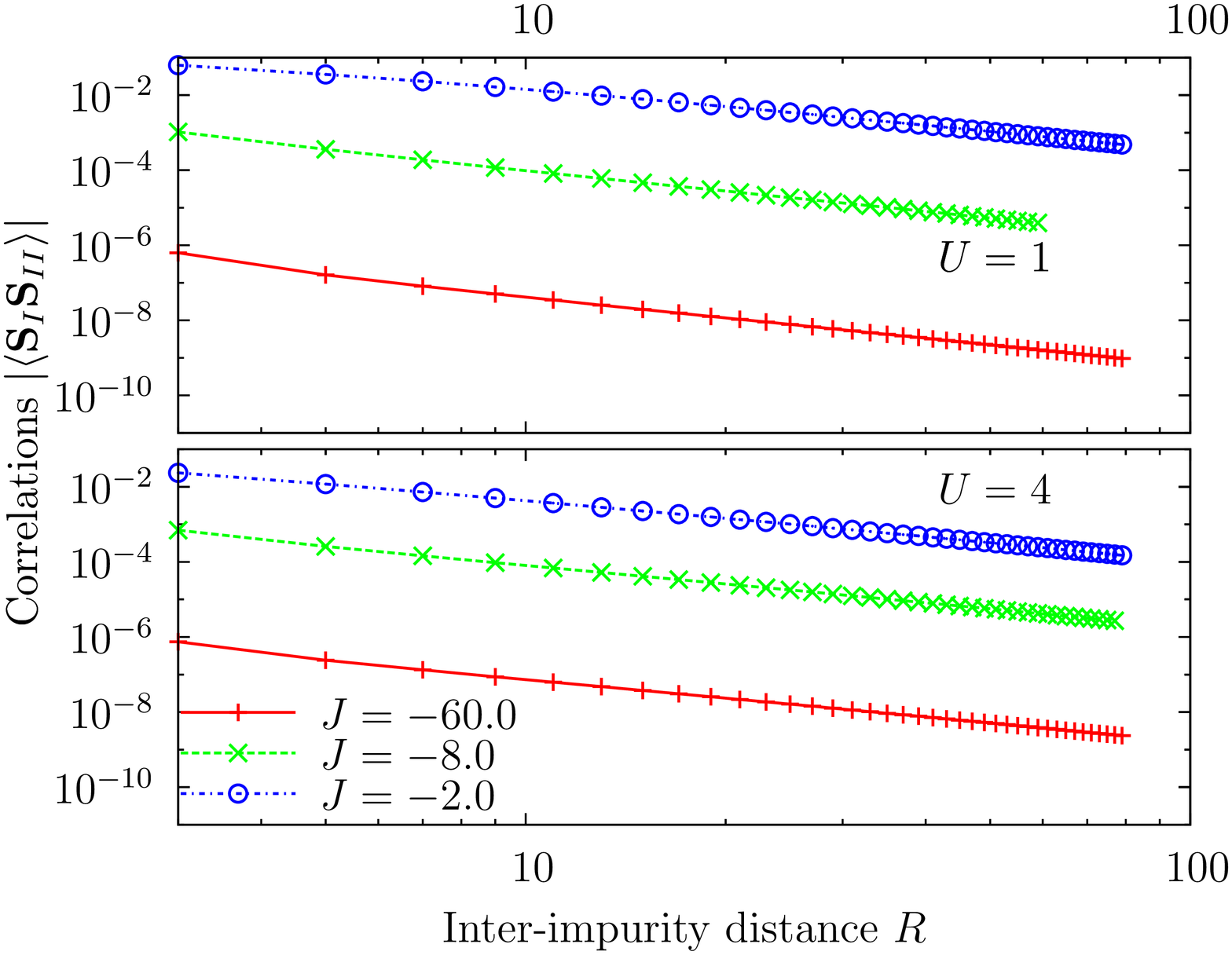}}
\caption{(Color online) DMRG data for $\langle \mathbf{S}_I \mathbf{S}_{II} \rangle_R$ on a double-logarithmic scale for several values of $J$  at $U=1$ as well as $U=4$ and the impurities attached to the ends.}
\label{fig: loglog_U14}
\end{figure}

\begin{figure}
\centering
\resizebox{0.98\columnwidth}{!}{\includegraphics[angle=-90]{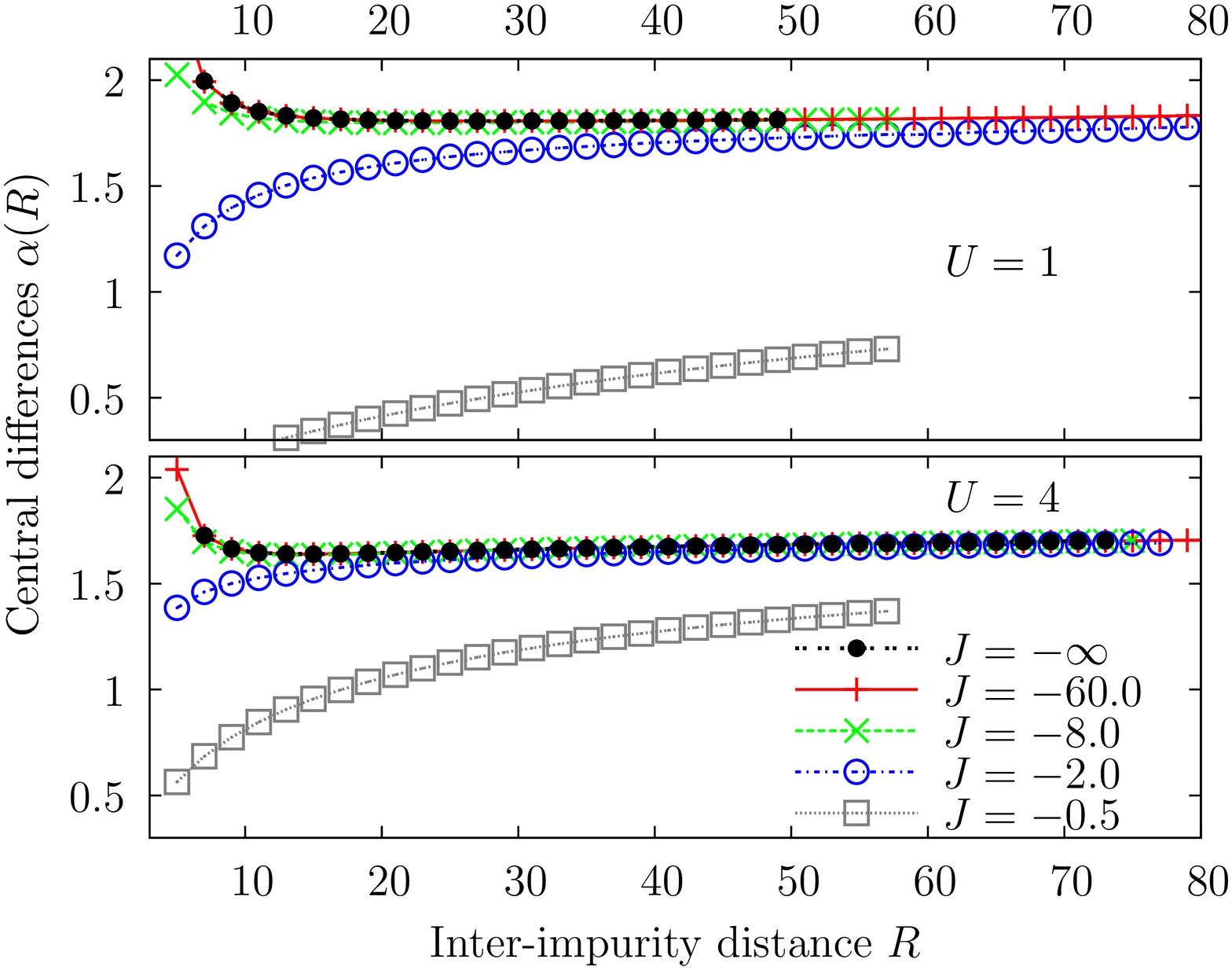}}
\caption{(Color online) Double-logarithmic central differences $\alpha(R)$ for $U=1$ as well as $U=4$ and various values of $J$ ($J=-\infty$ denoting the perturbative results). The impurities are attached to the ends of the chain.}
\label{fig: alpha_U14}
\end{figure}

\subsubsection{Interacting conduction chain $U\neq 0$}
We now switch on the interaction in the conduction chain. For $|J| \to \infty$, the system decouples into three parts as in the case of a noninteracting chain. Again, the impurity-impurity correlation function in this limit is proportional to the end-to-end correlation function between the two boundary sites of the interacting Hubbard chain between the Kondo singlets.
For the spin-spin correlations of the Hubbard chain, we can consult results from conformal field theory (CFT). For half-filling, a charge gap opens for all $U>0$ in the Hubbard chain, so that the system is a Mott insulator. The spin sector on the other hand remains gapless. In the continuum limit, the Hamiltonian of the Hubbard chain at half-filling separates into charge and spinon parts, both perturbed by marginal current-current interactions. From the spinon parts logarithmic corrections to the spin-spin correlation functions are expected due to the marginal irrelevant perturbation. However, on account of the SU(2) symmetry, these logarithmic corrections should have no $U$ dependence. From the holon parts we expect exponentially decaying correction terms $e^{-2MR}$ as two holons with mass $M$ are created. These terms are expected to decay much faster than the logarithmic correction terms. In summary, from conformal field theory for the continuum limit of the spin-spin correlation functions of the Hubbard 
chain with $U>0$, we still expect a $1/R^2$ decay with $U$-independent logarithmic corrections. However, it should be noted that these considerations have to be adapted to the special case of the end-to-end spin-spin correlations of an open Hubbard chain. 
 
Hallberg and Egger derived similar results from Luttinger liquid theory as the spin sector remains gapless and can therefore still be described by a Luttinger liquid with $g=0$.\cite{Hallberg1997} With bosonization, they calculated the  spin-spin correlations for the finite continuous system. By evaluating them close to the boundaries, they could show that for all values of $g$, the exponent of the end-to-end spin-spin correlation function asymptotically reaches $\alpha\rightarrow 2$, at that time in agreement with their DMRG data ($m=200$). Note, however, that from Fig.~\ref{fig: check_U1} we know that a too small value for $m$ can dramatically change the observed long-range behavior.

Let us therefore re-examine the long-distance behavior based on DMRG calculations with $m$ large enough to have a reasonably converged behavior at long distances. We start by analyzing our numerical data for a pure power-law decay given by Eq.~\eqref{eq:purepowerlaw}. Figure~\ref{fig: loglog_U14} shows the spin-spin correlation function $\langle \mathbf{S}_I \mathbf{S}_{II} \rangle_R$ on a double-logarithmic scale. Note that this representation suggests a very clear power law. The corresponding central differences extracted via Eq.~\eqref{eq: cdiff}, which give the exponent $\alpha$, are depicted in Fig.~\ref{fig: alpha_U14} for $U=1$ and $U=4$. The proportionality from Eq.~\eqref{eq: pt_universal} between $\langle \mathbf{S}_I \mathbf{S}_{II} \rangle^{(2)} $ and the correlations $\langle s^z_2  s^z_{L-1} \rangle$ in the conduction band has proven to be valid for nonzero values of $U$ as well. By calculating $\langle s^z_2  s^z_{L-1} \rangle$ for the corresponding open Hubbard chain of length $L-2$ via the 
DMRG method the asymptotic behavior of the central differences can be determined by the same perturbative approach. These data are labeled as $J=-\infty$ in Fig.~\ref{fig: alpha_U14}.  Overall, the DMRG data for nonzero $U$ suggest that the exponent $\alpha(R)$ approaches an asymptotic behavior for large couplings and distances, with the same difference between strong and weak coupling $J$ on how the asymptote is approached. The main result of our DMRG calculations, however, is the observation that the exponent assumes a value smaller than two. This result is, however, in clear contradiction to the expectations from CFT and bosonization.

\begin{figure}
\centering
\resizebox{0.98\columnwidth}{!}{\includegraphics[angle=-90]{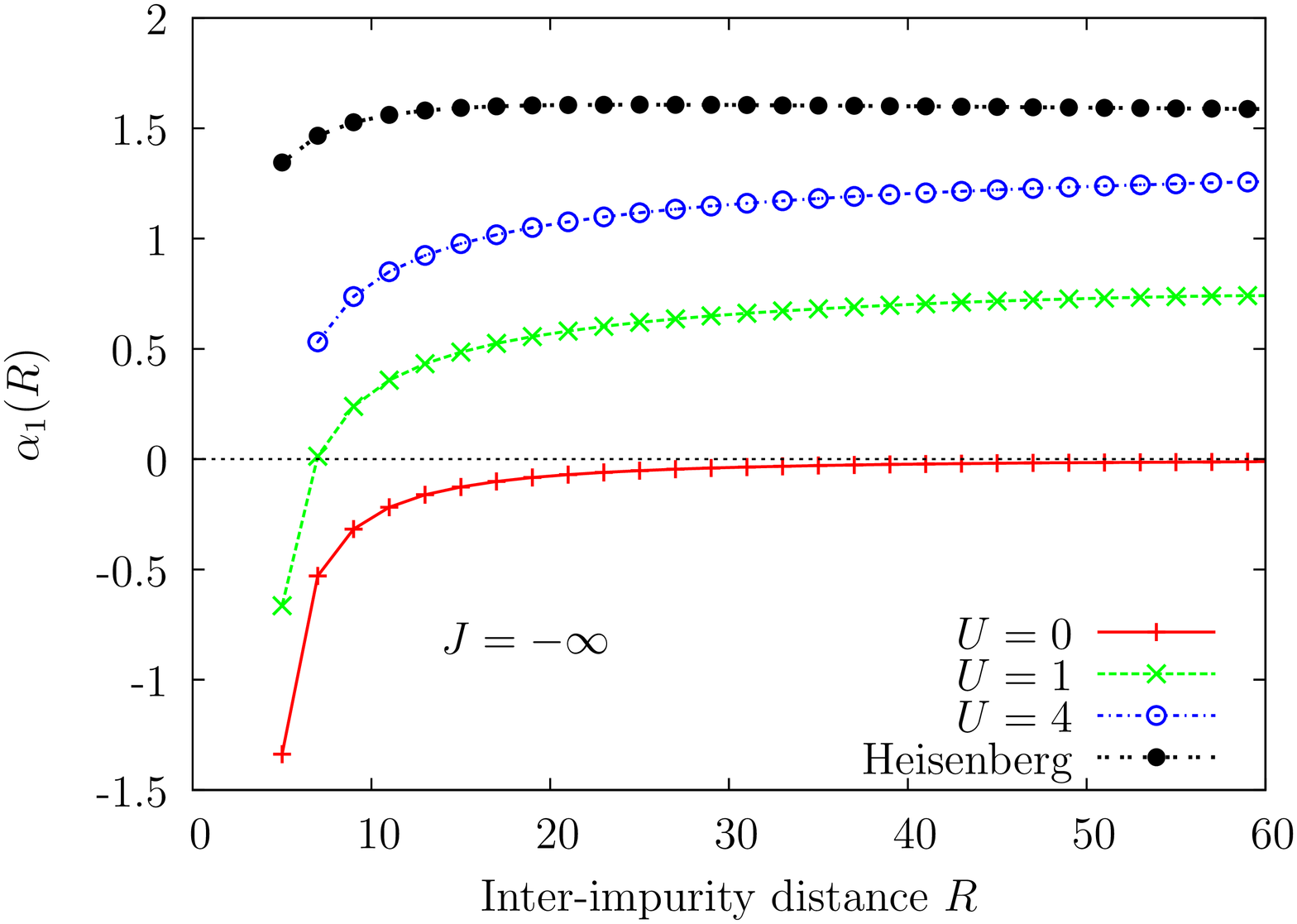}}
\caption{(Color online) Exponent $\alpha_1$ of the logarithmic correction as a function of the distance $R$ for $J=-\infty$ and several values of $U$.}
\label{fig: alpha1_vs_R}
\end{figure}

In order to resolve this contradiction, we now take into account the logarithmic correction appearing in the $2 k_F$ contribution to the correlation function, which is predicted by CFT as
\begin{align*}
    \langle \mathbf{S}_I \mathbf{S}_{II} \rangle \propto \, R^{-\alpha} \, \left[ \log(R)  \right]^{\alpha_1}.
\end{align*}
From this an expression for the exponent $\alpha_1$ of the logarithmic correction is derived via differentiation. One obtains
\begin{align*}
    \alpha_1 = \left( \frac{\mathrm{d} \log \langle \mathbf{S}_I \mathbf{S}_{II} \rangle}{\mathrm{d} \log R} + \alpha \right)\, \log(R),
\end{align*}
where we now use the exact result $\alpha=2$. The exponent $\alpha_1$ as a function of the distance $R$ is depicted in Fig.~\ref{fig: alpha1_vs_R}. As expected, $\alpha_1$ approaches zero rather rapidly for increasing distances and $U=0$. In the limit $U \to \infty$ the Heisenberg model gives $\alpha_1 \to 3/2$, which fits into the picture of an increasing exponent with $U$. For $U > 0$, $\alpha_1$ seems to converge to a finite value for increasing $R$, cf.\ Fig.~\ref{fig: alpha1_vs_R}. Note, however, that even for the largest systems studied here, one still finds an increasing value for $\alpha_1$. Furthermore, for fixed $U>0$, $\alpha_1$ seems to approach the same asymptotic value irrespective of $J$ (not shown here). In this sense, our DMRG data are compatible with an exponent $\alpha=2$ modified by a $U$-dependent logarithmic correction. However, the dependence on $U$ is not expected from conformal field theory. This contradiction will be discussed further in Sec.~\ref{sec: summary}.

\subsection{Impurities attached to the bulk}
\label{sec: bulk}

Next, we study the spatial dependence of $\langle \mathbf{S}_I \mathbf{S}_{II} \rangle$ in the bulk limit, i.e., $R \ll L/2$. For a noninteracting conduction chain in the large-$J$ limit, the system decomposes into five parts: a left and right conduction chain, the two singlets, and an effective chain between them. However, the spin-spin correlation functions between the two attached spins in this limit are again just proportional to the end-to-end spin-spin correlations between the two boundary sites of the chain in-between. The left and right chains show no dependence on the correlation functions in that limit. This was already argued by Hallberg and Egger.\cite{Hallberg1997} Consequently, the same expectations from conformal field theory and bosonization are valid as in the case of the impurities coupled to ends of the chain. 

Due to the oscillatory behavior of the correlation function, it is justified to merely consider odd inter-impurity distances $R$ in our DMRG calculations. Additionally, we choose even system sizes $L$ as this ensures a nondegenerate ground state. For $R=3,7,11,\dots$ the impurities are attached symmetrically around the center of the chain whereas for $R=5,9,13,\dots$ they are shifted out of this symmetric setup by one site. This provides us with the $L$-independent values of $\langle \mathbf{S}_I \mathbf{S}_{II} \rangle_R$ (cf.\ Appendix~\ref{sec: bulk_setup}). We worked with chains of length $L=120$ to make sure that the impurities are sufficiently far away from the ends to minimize boundary effects. 
 
\begin{figure}
\centering
\resizebox{0.98\columnwidth}{!}{\includegraphics[angle=-90]{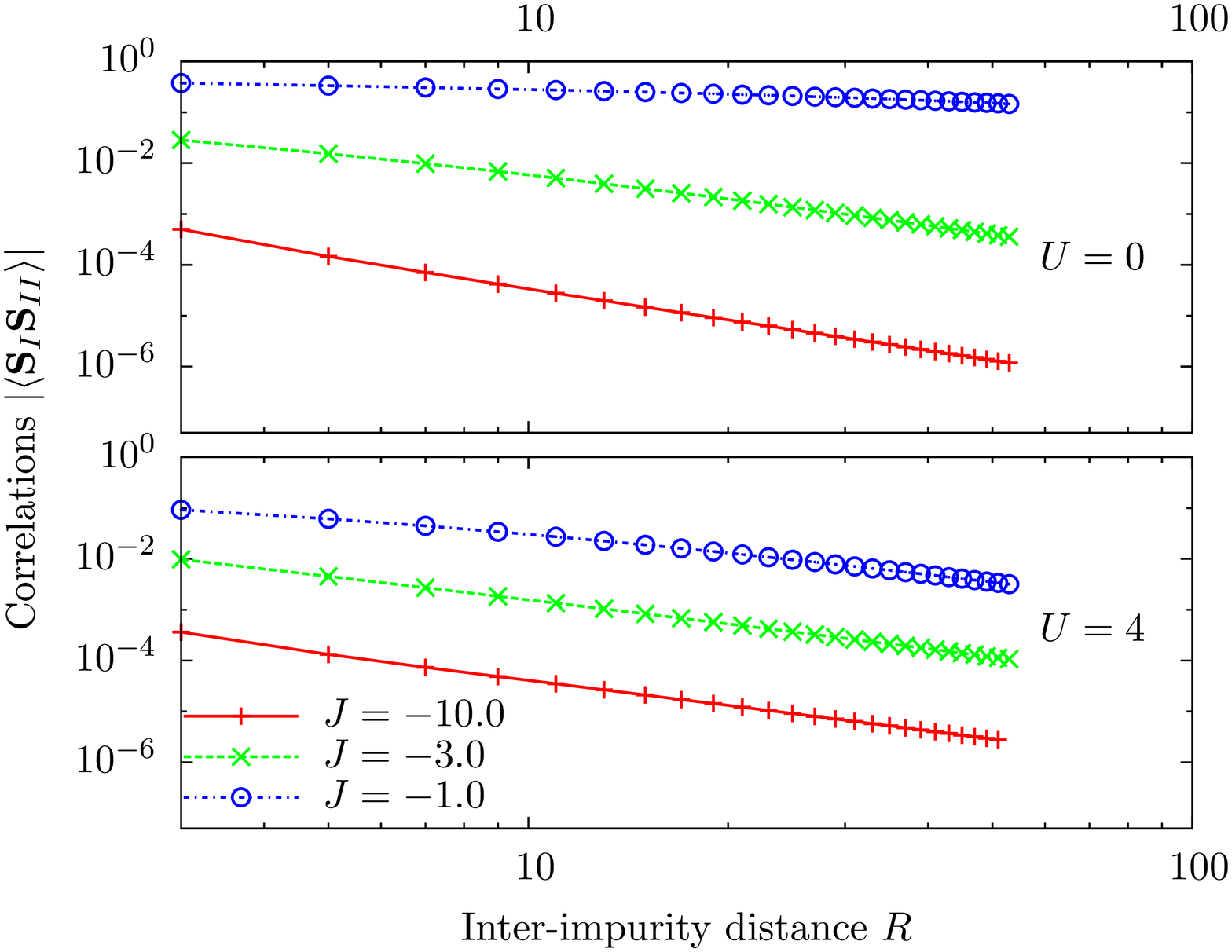}}
\caption{(Color online) DMRG data for $\langle \mathbf{S}_I \mathbf{S}_{II} \rangle_R$ on a double-logarithmic scale for several values of $J$  at $U=0$ as well as $U=4$ in the bulk limit.}
\label{fig: loglog_bulk_U04}
\end{figure}

\begin{figure}
\centering
\resizebox{0.98\columnwidth}{!}{\includegraphics[angle=-90]{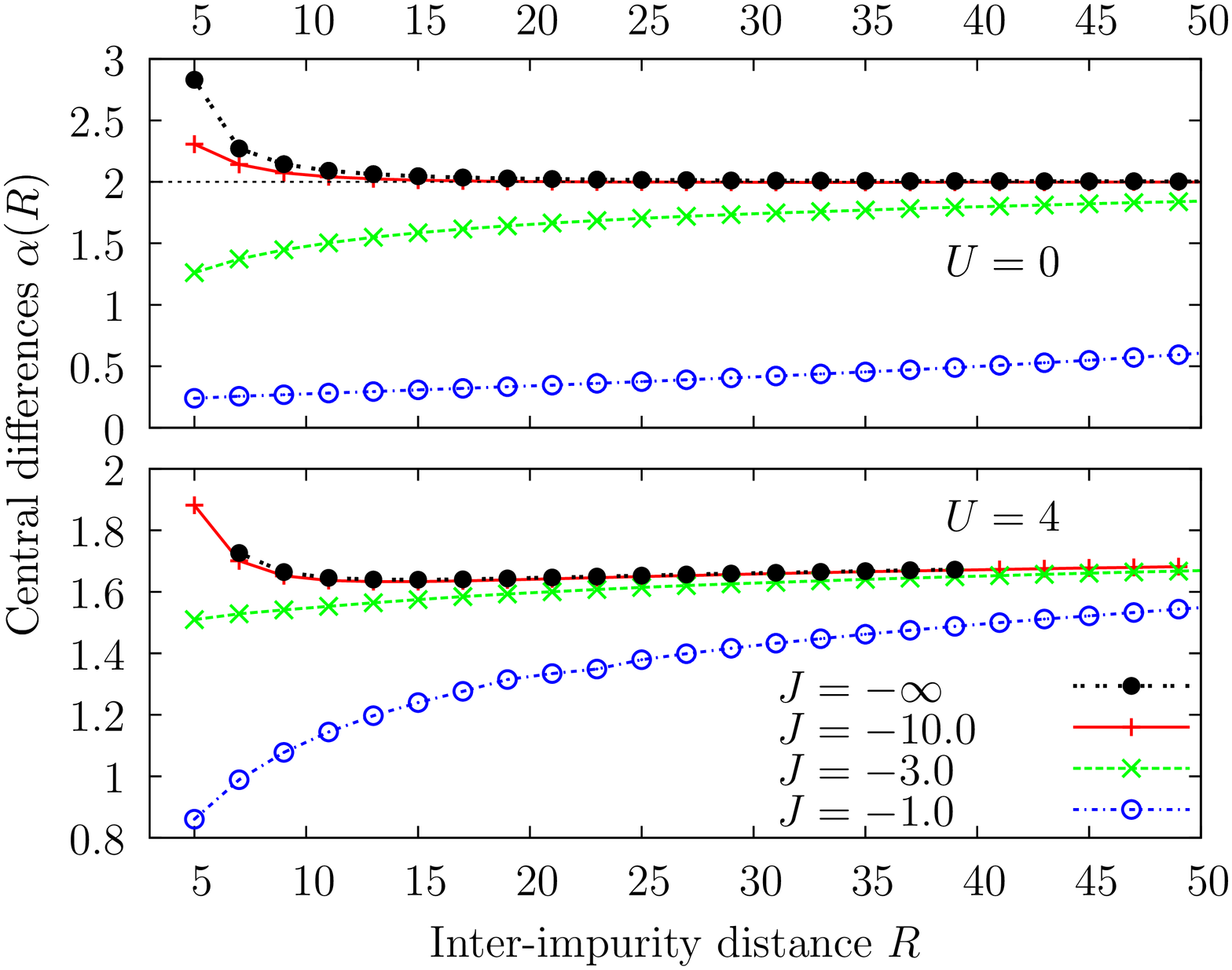}}
\caption{(Color online) Double-logarithmic central differences $\alpha(R)$ for $U=0$ and $U=4$ in the bulk limit for various couplings $J$ ($J=-\infty$ denoting the perturbative calculation).}
\label{fig: alpha_bulk_U04}
\end{figure}

Our DMRG results for the correlation functions between two impurities attached to the bulk are shown on a double-logarithmic scale in Fig.~\ref{fig: loglog_bulk_U04}. The qualitative behavior of $\langle \mathbf{S}_I \mathbf{S}_{II} \rangle_R$ is very similar to that observed for the impurities attached to the ends, cf. Figs.~\ref{fig: ptheory} and~\ref{fig: loglog_U14}. The upper panel of Fig.~\ref{fig: alpha_bulk_U04} contains an analysis by means of double-logarithmic differences for two impurities attached to the bulk of a noninteracting conduction chain. As for the system with the spins at the boundaries, both the large-$J$ and large-$R$ behavior obey a power law with an exponent of $\alpha \to 2$ for $U=0$. Moreover, the results are still consistent with perturbation theory. In the presence of correlations $U\neq 0$ in the bulk limit, we still expect the same decomposition of the system into five parts. Therefore, the asymptotic behavior for $U=4$ in the lower panel of Fig.~\ref{fig: alpha_bulk_U04} 
also turns out to be the same as for two impurities coupled to the ends in Fig.~\ref{fig: alpha_U14}. Consequently, if we assume a pure power law the bulk exponent assumes a value which is smaller than two as well.\\
 
In Fig.~\ref{fig: alpha_bulk_U04} it also has to be noted that the asymptotics are approached faster in the case of stronger correlations.\cite{Hallberg1997} This can be made plausible because the RKKY interaction, which is mediated via the spin polarization of the conduction band electrons by the magnetic moments of the impurities, is suppressed in the presence of stronger correlations. Thus, the crossover from RKKY to Kondo behavior occurs for smaller distances and couplings. 

\section{Summary and discussion}
\label{sec: summary}

Let us summarize our results for the long-distance behavior of the correlation function between the two impurities. We have numerically shown that the long-distance behavior for the two impurities in the bulk is equivalent to the long-distance behavior for two impurities coupled to the ends of the chain. This allows us to focus on the setup of one impurity coupled to either end of the chain~-- which is a computationally less demanding system.

It turns out that for a fixed value of $U$ the exponents of the power-law decay will converge to the same value for different $J<0$. The larger $|J|$, the better the convergence, which can be seen in Figs.~\ref{fig: alpha_U0} and \ref{fig: alpha_U14}. This has already been pointed out by Hallberg and Egger.\cite{Hallberg1997}  The last open question is whether the exponent depends on the interaction $U$ in the Hubbard chain.
In order to answer this question, we can focus on the large-$J$ limit. We have shown that our perturbation calculation agrees very well with our numerical data. For $U=0$, we find a value of $\alpha=2$ for the exponent which is in agreement with the result for the correlations of the noninteracting chain. For $U>0$, we also find seemingly converged exponents, but with $\alpha<2$. This is in contradiction to expectations from CFT and bosonization. In order to resolve this contradiction, we have analyzed the data for a $1/R^2$ decay with logarithmic corrections. Taking into account the logarithmic correction appearing in the $2 k_F$ contribution to the correlation function, we can show that our results are in agreement with a converged exponent $\alpha=2$, which would be consistent with CFT/bosonization. However, we obtain a $U$-dependent exponent $\alpha_1=\alpha_1(U)$ for the logarithmic correction, which again stands in contradiction to the results from CFT. 

The interpretation of this result is difficult. One has to remember that the CFT predictions, in particular the logarithmic corrections, are based on the continuum limit and do not include boundary effects. For the end-to-end spin correlations a $U$-dependent logarithmic correction may thus develop for finite systems if, due to the boundaries, a coupling between spin and charge degrees of freedom occurs. As one possible consequence, the $U$-dependent exponents of the logarithmic corrections may survive in the thermodynamic limit. Since the corrections are of logarithmic nature, we can infer that one naturally needs to access system sizes that are several orders larger. To our knowledge, such investigations have not been performed yet, but would be interesting in view of the results presented here. Alternatively, one may study a modified Hamiltonian in which the 
coupling to the marginal perturbation is reduced in order to suppress the logarithmic corrections.  This can be achieved, e.g., by adding an 
additional nearest-neighbor Coulomb interaction $V = - U /\left[2\,\cos(2k_F)\right]$ in the extended Hubbard model.\cite{PhysRevB.73.045125}

Furthermore, our observations urge us to be extremely careful in extracting exponents from numerical data. The double-logarithmic presentation of the correlation function suggests a clean power law and the extracted exponents look converged for the system sizes studied and suggest a value of $\alpha<2$ for $U>0$. However, we know that for much larger systems the true asymptotic value of $\alpha =2$ with logarithmic corrections will eventually be reached. A similar behavior has also been observed for the boundary exponent of the spectral function in open Hubbard chains. Here, for certain interactions in the studied systems of at most 500 sites, the power-law behavior with respect to the expected exponents did not occur at all.\cite{Meden2000,Schuricht2013,Soeffing2012} In the light of these findings, it is clear why our studies based on very accurate DMRG simulations lead to these strong deviations to the observation of the seemingly proper, $U$-independent exponent $\alpha=2$ by Hallberg and Egger. As their 
calculations were based on smaller distances with significantly less states kept, we can understand this discrepancy from the extreme sensitivity of the power-law exponents on the numerical accuracy. In order to fully resolve the long-distance behavior of correlations in the two-impurity Kondo-Hubbard chain, one would have to analyze system sizes that are much larger rendering further calculations using the DMRG impossible. In particular, the intrinsic exponential decay of long-range correlations in the DMRG makes these calculations of end-to-end correlation functions very expensive. Possible other algorithms to overcome this problem are, for example, MERA\cite{Vidal2007,Vidal2008,Evenbly2010} or other tensor networks, which are known to be able to recover the right form of the correlations.

\acknowledgments
We would like to acknowledge fruitful discussions with A. Honecker, S. Manmana and K.\ Sch\"onhammer. PED and TP further acknowledge the support by the DFG through the collaborative research center SFB 602. ACT thankfully appreciates the hospitality and cooperation at the Centro At\'omico Bariloche where a part of this work has been performed. Computer support by the GWDG and the GoeGrid project is also acknowledged.

\appendix
 
\section{Perturbation theory}
\label{sec: ptheory}

Let us first consider the case of a single impurity coupled to the first site of a noninteracting chain. In the strong-coupling limit $|J| \gg 1$, there is hardly any hopping between the sites 1 and 2 due to the rigidly bound singlet between the impurity and the spin at the first conduction band site. So, the corresponding hopping term represents the perturbation. The calculations in Ref.~\onlinecite{Sorensen1996} reveal that the expectation value to second order in $t$ of a single impurity spin can be stated as
\begin{align}
   \langle S^z_I \rangle^{(2)} = \frac{20}{9} \, \left( \frac{t}{J}\right)^2 \langle F_{L-1} | \psi_{2,\,\uparrow}^{\dagger} \psi_{2,\,\uparrow}^{\phantom{\dagger}} - \psi_{2,\,\downarrow}^{\dagger} \psi_{2,\,\downarrow}^{\phantom{\dagger}} | F_{L-1} \rangle, \label{eq: single_Imp}
\end{align}
where $|F_{L-1}\rangle$ is the ground state of a noninteracting chain of length $L-1$ and $\psi_{i,\,\sigma}^{(\dagger)}$ are fermionic field operators at site $i$. If one wants to introduce a second impurity at the other end of the chain, there is a slight modification as the first and the last conduction band sites will be quenched out. Thus, one uses $|F_{L-2}\rangle$ instead of $|F_{L-1}\rangle$ and also obtains an expression for $\langle S^z_{II} \rangle^{(2)}$, which is equivalent to Eq.~\eqref{eq: single_Imp} with $i=L-1$. Combining these two expressions yields the impurity-impurity correlation function 
\begin{align}
\langle S_{I}^z S_{II}^z  \rangle^{(2)}  &= 4 \left( \frac{20}{9} \right)^2 \left( \frac{t}{J} \right)^4 \langle F_{L-2} | s_2^z s_{L-1}^z |F_{L-2} \rangle \label{eq: SISII_combined}
\end{align}   
in the strong-coupling limit as the two singlets are screened from each other by the free chain of length $L-2$ in-between. Here $s_i^z = ( \psi_{i,\,\uparrow}^{\dagger} \psi_{i,\,\uparrow}^{\phantom{\dagger}} - \psi_{i,\,\downarrow}^{\dagger} \psi_{i,\,\downarrow}^{\phantom{\dagger}} )/2$ denote the spin operators. The correlations $\langle s_2^z s_{L-1}^z \rangle$ can be calculated analytically for a noninteracting chain at half-filling and for $S^z_{tot} =0$. Inserting the explicit representation of the field operators
\begin{align*}
\psi_{j,\,\sigma}^{(\dagger)} = \sqrt{\frac{2}{L+1}} \sum_{ \{k_n\}} \sin(k_n j)\,c_{k_n}^{(\dagger)},\label{eq: feldop}
\end{align*}
where $k_n = n\pi/(L+1)$ and $n=1,\dots,L$, gives 
\begin{align*}
  \langle F_L | &s_2^z s_{L-1}^z | F_L \rangle = \frac{2}{(L+1)^2} \sum_{l=1}^{L/2} \sum_{q=L/2+1}^{L} \biggl[ \sin \left( \frac{L-1}{L+1}\pi l \right) \times \\
 &\times \sin \left( \frac{2\pi}{L+1} l \right)  \sin \left( \frac{L-1}{L+1}\pi q \right) \sin \left( \frac{2\pi}{L+1} q \right) \biggr].
\end{align*}
Now Eq.~\eqref{eq: ptheory} directly follows upon exploiting this result, Eq.~\eqref{eq: SISII_combined} and the rotational invariance of the system.

\section{Bulk setup}
\label{sec: bulk_setup}

\begin{figure}[htbp!]
\centering
\resizebox{0.98\columnwidth}{!}{\includegraphics[angle=-90]{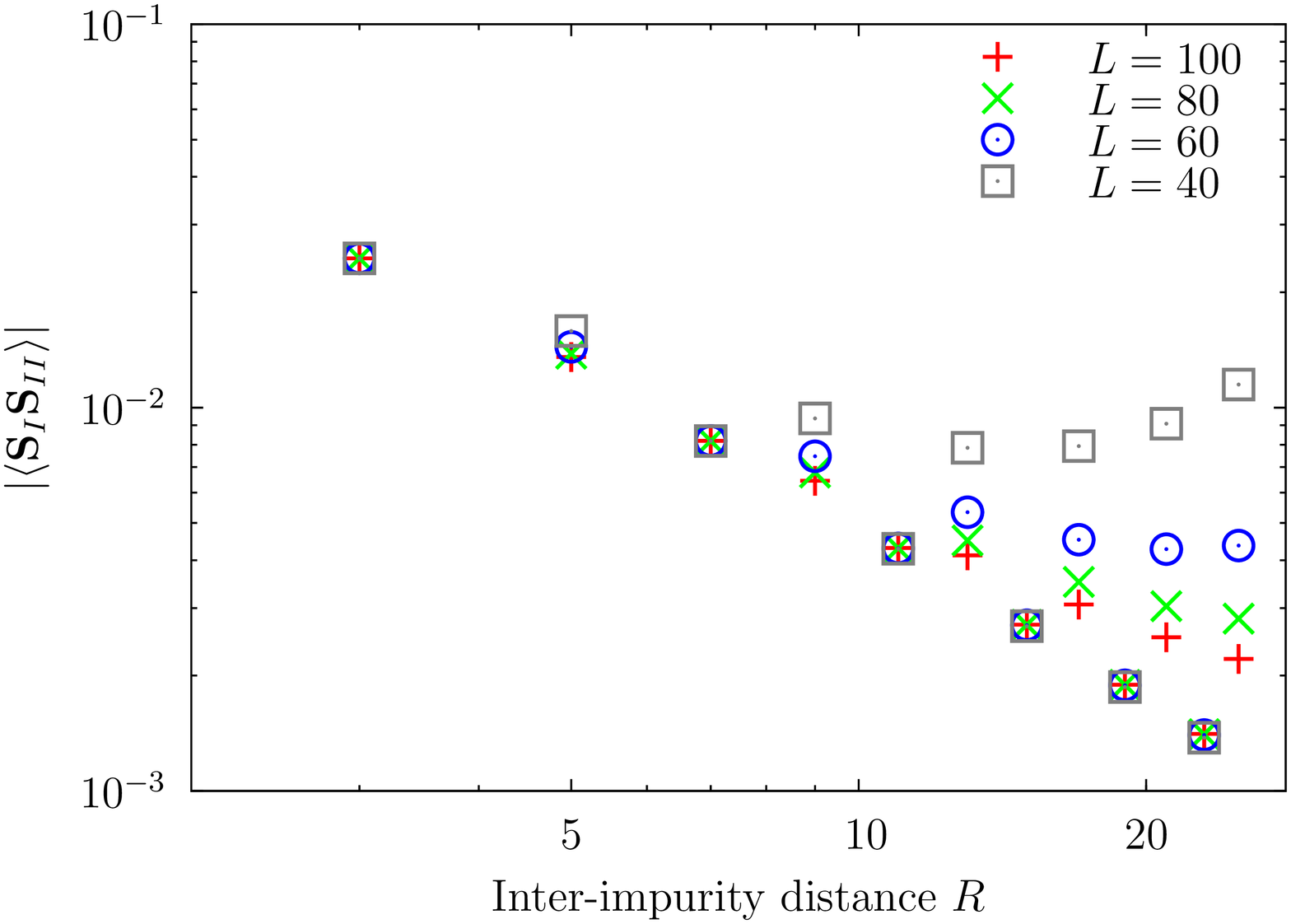}}
\caption{(Color online) Finite-size effects in the bulk limit ($R \ll L/2$) for impurities attached symmetrically around the center of the chain for $U=4$ and $J=-2$.}
\label{fig: FSE}
\end{figure}

As described in Sec.~\ref{sec: bulk}, we regard odd distances $R$ and even chains of length $L$ to ensure a nondegenerate ground state. If one attaches the impurities symmetrically around the center of the chain, the correlation function will have nearly the same value for $(R+1)/2$ even irrespective of $L$ as long as  $R \ll L/2$ is fulfilled, while it will have a strong size dependence for other odd values of $R$. This is shown for $U=4$ and $J=-2$ in Fig.~\ref{fig: FSE}. By means of a finite-size scaling with respect to $1/L$ the extrapolated values for the other distances $R=5,9,13,\dots$ will obey the same behavior for $L \to \infty $ as the $L$-independent $R$'s.

\bibliography{References}	

\end{document}